\documentclass[prl,aps,twocolumn,showpacs,preprintnumbers,amsmath,superscriptaddress,floatfix,10pt]{revtex4}

\usepackage{amsmath}
\usepackage[latin1]{inputenc}
\usepackage{bm}
\usepackage{graphicx}
\usepackage{SIunits}
\usepackage{subfigure}

\newcommand{\be}{\begin{equation}}
\newcommand{\ee}{\end{equation}}
\newcommand{\ba}{\begin{eqnarray}}
\newcommand{\ea}{\end{eqnarray}}
\newcommand{\ban}{\begin{eqnarray*}}
\newcommand{\ean}{\end{eqnarray*}}

\begin{document}

\title{\large{Nanosecond-scale timing jitter in transition edge sensors at telecom and visible wavelengths}}

\author{Antia Lamas-Linares}
\affiliation{National Institute of Standards and Technology, 325 Broadway St., Boulder CO 80305, USA}
\author{Brice Calkins}
\affiliation{National Institute of Standards and Technology, 325 Broadway St., Boulder CO 80305, USA}
\author{Nathan A. Tomlin}
\affiliation{National Institute of Standards and Technology, 325 Broadway St., Boulder CO 80305, USA}
\author{Thomas Gerrits}
\affiliation{National Institute of Standards and Technology, 325 Broadway St., Boulder CO 80305, USA}
\author{Adriana E. Lita}
\affiliation{National Institute of Standards and Technology, 325 Broadway St., Boulder CO 80305, USA}
\author{J\"{o}rn Beyer}
\affiliation{Phys.--Tech. Bundesanstalt (PTB), Berlin, Germany}
\author{Richard P. Mirin}
\affiliation{National Institute of Standards and Technology, 325 Broadway St., Boulder CO 80305, USA}
\author{Sae Woo Nam}
\affiliation{National Institute of Standards and Technology, 325 Broadway St., Boulder CO 80305, USA}
\date{\today}

\begin{abstract}
Transition edge sensors (TES) have the highest reported efficiencies ($>98\%$) for detection of single photons in the visible and near infrared. Experiments in quantum information and foundations of physics that rely critically on this efficiency have started incorporating these detectors into conventional quantum optics setups. However, their range of applicability has been hindered by slow operation both in recovery time and timing jitter. We show here how a conventional tungsten-TES can be operated with jitter times of $\leq 4\;{\rm ns}$, well within the timing resolution necessary for MHz clocking of experiments, and providing an important practical simplification for experiments that rely on the simultaneous closing of both efficiency and locality loopholes.
\end{abstract}


\maketitle

The last few years have yielded impressive advances in the technology of single photon detectors. Of particular relevance are the cryogenic detectors based on superconducting nanowires and on superconducting transition edge sensors (TES). The former have shown very small timing jitter distributions, as low as $30\;{\rm ps}$, and quantum efficiency (QE) values recently reported in the $90\%$ range~\cite{baek:11,verma:12}. The latter have the highest reported single photon detection efficiencies (up to 98\%)\cite{lita:10,fukuda:11b} and are inherently photon number resolving\cite{lita:08,fukuda:11}, qualities that are critical for experiments in quantum information science and quantum optics.

A wider applicability of TES in quantum optics experiments has been limited by their relative slowness, both in terms of recovery and jitter times. Recovery times on the order of microseconds are the norm\cite{eisaman:11} and limit or complicate high photon flux detection. Commonly reported timing uncertainties on the order of $100\;{\rm ns}$ pose even more of a problem, since they severely constrain the experiments that could benefit most from the high efficiencies and photon number resolving capabilities: loophole free Bell inequalities and multi-photon entanglement generation with short pulsed lasers. For loophole free Bell inequalities, the large coincidence windows required by the large jitter impose longer distances between detectors in order to close the locality condition, which in turn compromise the system detection efficiency. Multi-photon entangled states are very often produced with Ti:Sapph mode-locked lasers at a repetition rate of $80\;{\rm MHz}$. Even the best reported jitter values ($28\;{\rm ns}$)\cite{fukuda:11} are not compatible with these moderate repetition rates, and are much larger than what can be obtained by commercially available Si single photon avalanche photo diodes\cite{eisaman:11}.

In this letter we demonstrate detection of $1550\;{\rm nm}$ photons with jitter values of $\leq\;4\;{\rm ns}$ for a tungsten-TES (W-TES) read out with a low input inductance SQUID (superconducting quantum interference device) amplifier. Operation in this low jitter regime retains the existing qualities of photon number resolution and high quantum efficiency. The TES films used for this test have been fabricated for optimum quantum efficiency and photon number resolution (as previously described\cite{lita:10}) with no special steps taken to optimize the speed of the device itself.
 
\begin{figure}  \includegraphics[width=3.5in]{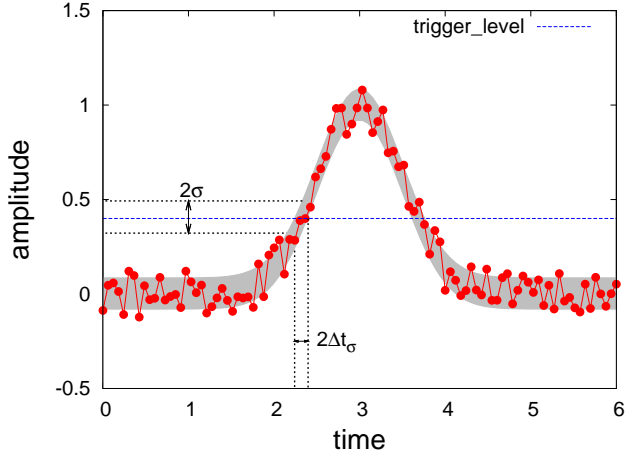}
\caption{Jitter dependence on noise and slope for a simulated Gaussian pulse with random noise. The timing uncertainty, $\Delta t$, of a pulse on crossing a threshold level is a function of the noise on the signal, $\sigma$, and the local slope of the pulse at the threshold crossing point.}
\label{fig:jitter_cartoon}
\end{figure}

For any timing pulse signal, the jitter or timing uncertainty for crossing a threshold is determined by the noise and the underlying slope of the signal at the point of crossing (see fig.~\ref{fig:jitter_cartoon}),
\begin{equation}
\Delta t_{\sigma}=\frac{\sigma}{\frac{dA}{dt}\big |_t} \approx \frac{\sigma}{A_{\rm max}}\tau_{\rm rise}, \label{eq:jitter_gen}
\end{equation}
where $A$ represents the amplitude of the signal, $\sigma$ its standard deviation and $\tau_{\rm rise}$ the rise time. The approximation in equation~\ref{eq:jitter_gen} holds for a linear rise of the pulse, and we will use this expression as a guide to the expected jitter performance. For a TES, we can calculate the expected RMS (root mean square) noise and amplitude of the current signal produced by the arrival of a photon~\cite{calkins:12}. Assuming instant thermalization, no damping inductance and for an ideal voltage bias, the change in current is given~\cite{irwin:05} by
\begin{equation}
\Delta I \approx \sqrt{\frac{P_0}{R_0}}\frac{\alpha\eta h \nu}{C T_0 (1+\beta)} ,
\end{equation}
where $P_0$ is the equilibrium power dissipation of the device, $R_0$ is the resistance of the device at the operating point, $C$ is the device heat capacity, $\eta$ is the energy collection fraction, $T_0$ and $I_0$ are the temperature and current at the operation point, $\alpha = \frac{T_0}{R_0}\frac{\partial R}{\partial T}$ and $\beta = \frac{I_0}{R_0}\frac{\partial R}{\partial I}$ are related to the shape of the superconducting to normal transition, and $h \nu$ is the absorbed photon energy. 

The main contribution to the noise in the TES signal is a combination of Johnson noise in the device and thermal fluctuations between the device and the bath,
\begin{equation}
I^{2}_{RMS} \approx \frac{\sqrt{2}k_BT_0(1+2\beta)(1+M^{2}_J)}{L(1+\beta)},
\end{equation}
where $M_J$ is a phenomenological parameter representing the excess Johnson noise, and with the implicit assumption that the SQUID contribution to the noise is sufficiently small. In the limit of the device recovery time being much longer than the rise time, the intrinsic rise time, $\tau_{{\rm el}}$, of the photon detection pulse is given by the inductance and resistance of the TES-SQUID combination,
\begin{equation}
\tau_{{\rm el}}=\frac{L}{R_0(1+\beta)},
\end{equation}
where $L$ is the input inductance of the SQUID amplifier including wiring and parasitic contributions. Also, the bandwidth of the external (in this case, room temperature) amplifier ($\Delta f$), will limit the performance, giving a combined rise time of
\begin{equation}
\tau_{{\rm rise}}=\sqrt{\tau_{{\rm el}}^{2}+\tau_{{\rm ext}}^{2}},
\end{equation}
where the rise time of the external amplifier is related to the bandwidth as $\tau_{{\rm ext}}\approx 0.35/\Delta f$. Combining these expressions and with the additional simplifying assumptions of a noiseless amplifier, operation at the superconducting transition temperature and low base temperature, we arrive at a final expression for the jitter~\cite{calkins:12},
\begin{eqnarray}
\Delta t_{FWHM} &\approx& \frac{2\sqrt{2 \ln 2} I_{RMS}\;\tau_{{\rm rise}}}{\Delta I} \nonumber \\ 
&=& \sqrt {8 \ln 2 \sqrt{2} (1+\beta)(1+2 \beta)(1+M_{J}^2)} \nonumber \\ 
&\times& \frac{\gamma}{\alpha \eta h \nu} \times \sqrt{\frac{R_0 V k_B}{L \Sigma}} \label{eq:jitter} \\ \nonumber
&\times& \sqrt{\tau_{\rm ext}^{2}+\frac{L^2}{R_{0}^2(1+\beta)^2}},
\end{eqnarray}
where $\Sigma$ and $\gamma$ are material parameters and $V$ is the volume.

The device under test is a W-TES optimized for detection of near-IR photons at a wavelength of approximately $800\;{\rm nm}$. The physical parameters and material characteristics of this particular TES-SQUID system are listed in Table~\ref{tab:param}.
\begin{table}
\caption{Material parameters in TES device. Some quantities are shown as a range, reflecting our limited knowledge of the exact values.}
\centering
\begin{tabular}{|l|r||l|r|}
\hline
$T_0$&$150\;{\rm mK}$ & $h\nu$ & $0.8 {\rm \;eV}$\\
\hline
$R_0$&$1\;\Omega$& $\tau_{ext}$&$17.5\;{\rm ns}$\\
\hline
$V$ & $12.5\;{\rm \mu m}^3$& $M_J$& 1.5-3.5 \\
\hline
$\Sigma$ & $0.4\;{\rm nW}\mu {\rm m}^{-3} {\rm K}^{-1}$& $L$& $24\pm5\;{\rm nH}$\\
\hline
$\alpha$ & 150-800&$\gamma$ & $340.2\;{\rm aJ\mu m^{-3} K^{-1}}$\\
\hline
$\beta$ & 0.8-2.2&$\eta$ & 0.4-0.9\\
\hline
\end{tabular}
\label{tab:param}
\end{table}

We measure the W-TES in a dilution refrigerator at a base temperature of $30\;{\rm mK}$. Critically to the timing performance, the TES is electrically connected to a low input inductance SQUID amplifier\cite{drung:07} via Al bond wires. Room temperature electronics perform the last amplification stage at a nominal bandwidth of $20\;{\rm MHz}$. The test light signal consists of a pulsed diode laser at a wavelength of $1550\;{\rm nm}$, a pulse duration of $1\;{\rm ns}$ and the repetition rate is kept at $100\;{\rm kHz}$, well below the recovery time of the TES, for convenience in the analysis. The signal was captured by a digitizing oscilloscope at a sampling rate of 1.25GS/s and 8 bit dynamic range. We chose the input optical power level to see a significant number of 1, 2, and 3 photon events to allow independent jitter analysis for different photon numbers. The SQUID and W-TES bias point were chosen manually to minimize rise time and maximize amplitude of the pulses.
\begin{figure}  \includegraphics[width=3.5in]{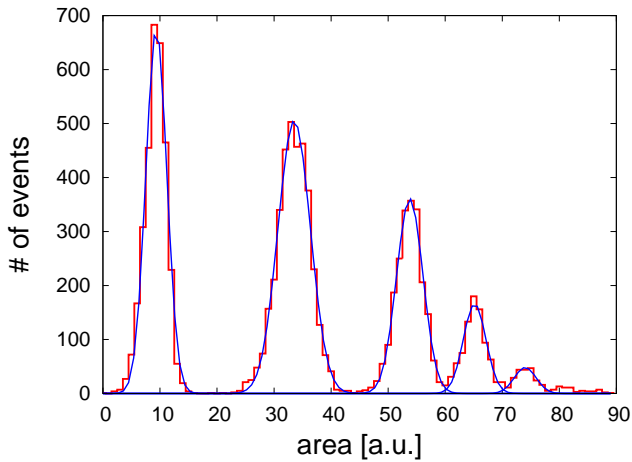}
\caption{Photon number resolution of device under test. The photon number peaks for 0, 1, 2, 3, 4 photons are clearly resolved. The inter-peak separation is highly non-linear as a consequence of an open loop operation of the squid amplification process and the non-linearity of the TES response.}
\label{fig:pnr}
\end{figure}

Figure~\ref{fig:pnr} shows a histogram of the areas of each pulse after a postprocessing digital matched filter. The peaks corresponding to 0, 1, 2, 3 and 4 photons are clearly visible, as is a strong non-linearity in the peak separation between consecutive photon numbers. This non-linearity is dominated by the SQUID response in open-loop operation, which we use for optimum bandwidth. The non-linear response complicates the determination of the energy resolution, as it varies strongly depending on the initial biasing point of the SQUID and with the energy dependent perturbation of this point. A postprocessing linearization of the energy scale using the known wavelength of the light used for testing, allows us to estimate the energy resolution to be $0.33 \pm 0.02\;{\rm eV}$, slightly worse than our typical value of $0.25\;{\rm eV}$. This linearization for the purpose of determining energy resolution is independent of, and has no influence on, the timing jitter results.


\begin{figure}  \includegraphics[width=3.5in]{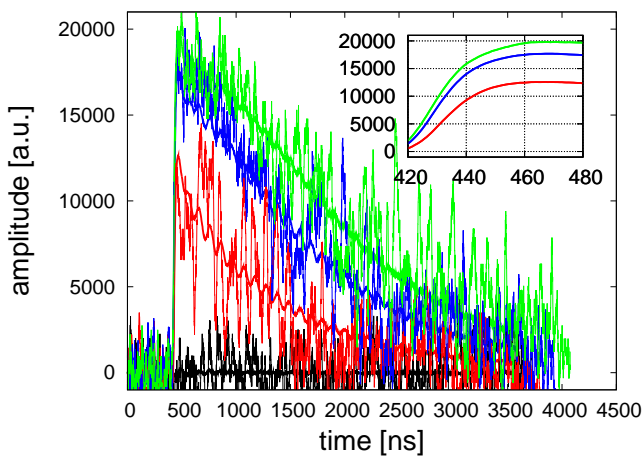}
\caption{Photon detection traces. The average pulse shape and a representative single shot trace for 1, 2 and 3 photons (of increasing heights) shows a short rise time in a pulse with a total duration of several microseconds. The 10\% to 90\% time is $24.8\;{\rm ns}$, $25.6\;{\rm ns}$ and $27.2\;{\rm ns}$ for the 1, 2 and 3 photon traces respectively. Decay times ($1/e$) are $759\;{\rm ns}$, $1278\;{\rm ns}$ and $1692\;{\rm ns}$. The inset shows the rise section of the photon detection average signals.}
\label{fig:ave_traces}
\end{figure}
The average pulse shapes for 1, 2 and 3 photon pulses are shown in Figure~\ref{fig:ave_traces.eps}. The 10\%-90\% rise time of these are $24.8\;{\rm ns}$, $25.6\;{\rm ns}$ and $27.2\;{\rm ns}$ respectively for the 1, 2 and 3 photon pulses. The corresponding $1/e$ decay times are $759\;{\rm ns}$, $1278\;{\rm ns}$ and $1692\;{\rm ns}$. The unequal decay times can be attributed to operating the TES device in an open-loop configuration; the non-linearity of the SQUID in this mode compresses the signal height for the larger photon number signals. Oscillations in the signal are possibly caused by ringing in the amplifier/SQUID/wiring system but should not significantly affect the conclusions from the data analysis.

\begin{figure}  \includegraphics[width=3.5in]{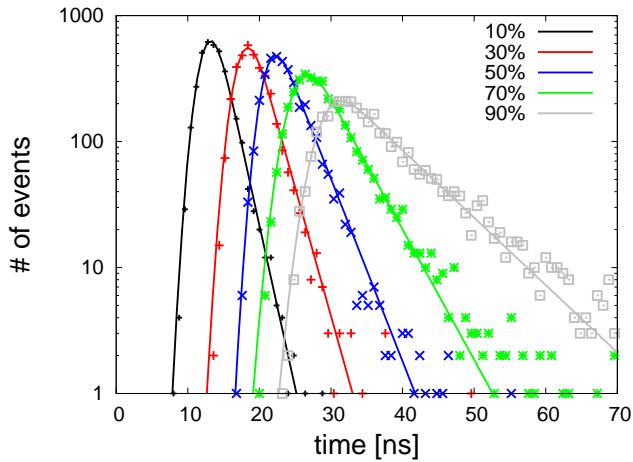}
\caption{Timing uncertainty. The histogram of crossing times is fitted to a convolution of a Gaussian and an exponential decay. The figure shows an example of these fits for 1-photon signals at threshold levels of $10\%$, $30\%$, $50\%$, $70\%$ and $90\%$. The fits for other threshold levels and for the 2 and 3 photon signals are of similar quality.}
\label{fig:conv_fit}
\end{figure}
\begin{figure}  \includegraphics[width=3.5in]{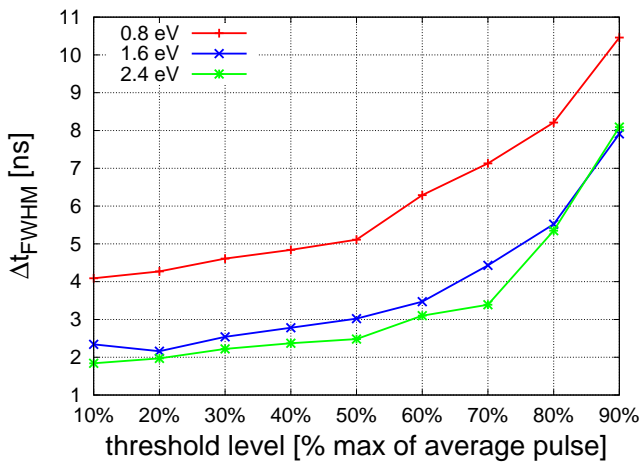}
\caption{Timing uncertainty. Thresholding at different levels provides the simplest method for measuring the time of arrival of a pulse and its associated uncertainty. The three curves show results for 1, 2 and 3 photon pulses ($0.8\;{\rm eV}$, $1.6\;{\rm eV}$ and $2.4\;{\rm eV}$ respectively).}
\label{fig:thresholdsigmas}
\end{figure}

We analyze the timing performance by first separating the signals according to the photon number (or energy) and subsequently applying a filter to determine the time of arrival of the pulse to each group independently. The simplest way to determine pulse timing is by setting a threshold for the signal at some fixed level which we take as a fraction of the maximum of the average pulse. Fitting the histogram of the crossing times to the convolution of a Gaussian and an exponential decay, 
\begin{eqnarray}
g(t) &=& A e^{\frac{-(t-t_0)^2}{2 \sigma^2}}\nonumber\\ 
d(t) &=& u(t) e^{-\tau t}\nonumber\\
f(t)&=&A \sigma {\sqrt \frac{\pi}{2}} e^{\frac{1}{2}\tau^2\sigma^2}e^{-\tau(t-t_0)}\\
&\times&{\rm erfc} \left[\frac{-t+t_0+\tau\sigma^2}{{\sqrt 2}\sigma}\right ]\nonumber,
\end{eqnarray}
provides our value for the jitter (see Figure~\ref{fig:conv_fit} for some fit examples). The variation of the FWHM times as a function of chosen threshold level is shown in figure~\ref{fig:thresholdsigmas}. As expected from eq.~\ref{eq:jitter_gen}, the jitter improves with the local steepness of the signal, which in the case of TES pulses is right at the onset, i.e. at lower threshold levels. For $1550\;{\rm nm}$ single photons, the fitted FWHM values of the timing uncertainty vary between $4.1\;{\rm ns}$ and $10.5\;{\rm ns}$. For the 2-photon signal, or equivalently for single photons at $775\;{\rm nm}$, the times are between $2.3\;{\rm ns}$ and $7.9\;{\rm ns}$, and become shorter as higher photon signals are considered. In all cases, the jitter is well within the $12\;{\rm ns}$ limiting case for operation with an $80\;{\rm MHz}$ repetition laser.

\begin{figure}  \includegraphics[width=3.5in]{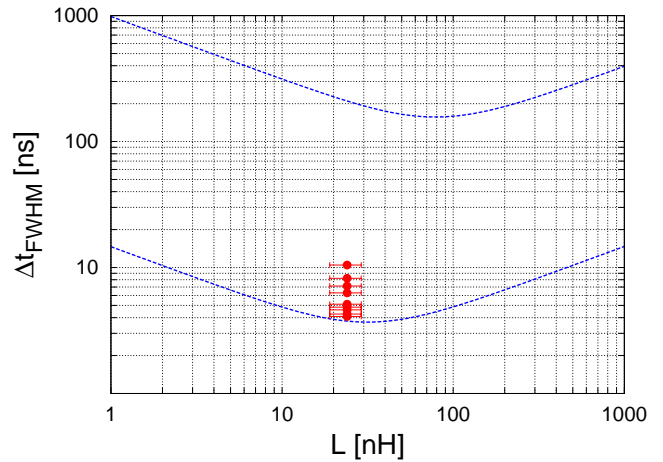}
\caption{Expected jitter for $1550\;{\rm nm}$ photons. The two extremal curves are the highest and lowest values of jitter calculated for the range of parameters values in our devices.  The data points correspond to the measured FWHM of the timing distributions for different trigger levels. Note that our measured value for the effective inductance ($L=24\;{\rm nH}$), is close to optimal.}
\label{fig:jittertheory}
\end{figure}

These numbers are roughly consistent with what is expected from eq.~\ref{eq:jitter} and the TES-SQUID parameters. In particular, the SQUID input noise~\cite{drung:07} needs to be less than $\approx 1\;{\rm pA/\sqrt{Hz}}$ for our assumption of negligible SQUID noise contribution and the resulting expression to be valid. Some of the physical and material parameters listed in table~\ref{tab:param} are known within a relatively small error. However, the parameters associated with the shape of the transition, $\alpha$ and $\beta$, and excess noise terms, $M_J$, are only approximately known for this device. Given these constraints, eq.~\ref{eq:jitter} predicts a wide range of possible values of $\Delta t_{FWHM}=3.9\;{\rm ns}-227\;{\rm ns}$. It is worth noting that, in the regime where the room temperature amplifier limits the rise time, there is an optimal choice of the term $\frac{L}{R_0(1+\beta)}=\tau_{{\rm el}}$. In our case, we believe our measured value of $L=24\;{\rm nH}$ with our adjustment of the operating point is near optimal for this system (see fig.~\ref{fig:jittertheory}).

In conclusion, optical TES have demonstrated extremely good photon number discrimination and close to unity quantum efficiency. However, they lag behind other detector technologies in their timing performance, both in the jitter or ``time of arrival'' and in their recovery times. In this paper we have shown how the jitter in these devices can be made significantly smaller than is ordinarily reported by using reduced input inductance SQUID amplifiers~\cite{drung:07}, achieving values as low as $4.1\;{\rm ns}$ for $1550\;{\rm nm}$ single photons and $2.3\;{\rm ns}$ for $775\;{\rm nm}$. These values are an order of magnitude smaller than previously reported and well below the technologically important $12\;{\rm ns}$ threshold associated with $80\;{\rm MHz}$ repetition Ti:Sa lasers.

We acknowledge Burm Baek, Marty Stevens and John Lehman for useful discussions and help with critical pieces of equipment. This work is partially funded by the NIST Quantum Information Initiative and a NIST-ARRA fellowship.

Contribution of NIST, an agency of the U.S. government, not subject to copyright.

\end{document}